\documentclass[api,prl,showpacs,letterpaper,twocolumn,groupedaddress,footinbib,amssymb]{revtex4}
\setcitestyle{super}
\usepackage{graphicx}
\usepackage[usenames]{color}
\usepackage{amstext}
\usepackage{amsmath}
\usepackage{amssymb}
\usepackage{amsfonts}

\begin{document}

\title{Topologically Protected Magnetic Helix for All-Spin-Based Applications}

\author{E. Y. Vedmedenko$^{1}$}
\email[corresp.\ author: ]{vedmeden@physnet.uni-hamburg.de}
\author{D. Altwein$^{1}$}
\affiliation{$^{1}$University of Hamburg, Institute for Applied Physics, Jungiusstr. 11a, 20355 Hamburg}

\date{\today}

\begin{abstract}
The recent years have witnessed an emergence of the field of all-spin-based devices without any flow of charge. An ultimate goal of this scientific direction is the realization of full spectrum of spin-based networks like in modern electronics. The concepts of energy storing elements, indispensable for those networks, are so far lacking. Analyzing analytically the size dependent properties of magnetic chains that are coupled via either exchange or long-range dipolar or Ruderman-Kittel-Kasuya-Yosida interactions, we discover a particularly simple law: magnetic configurations corresponding to helices with integer number of twists, that are commensurate with the chain's length, are energetically stable. This finding, supported by simulations and an experimentally benchmarked model, agrees with the study [J. Appl. Phys. \textbf{111}, 07E116 (2012)] showing that boundaries can topologically stabilize structures that are not stable otherwise. On that basis an energy storing element is proposed.
\end{abstract}
\maketitle

A challenge for the solid state physics at the nanoscale is to develop energetically efficient information and communication technologies. While spintronics, spin caloritronics and magnonics focus on the interaction of spins with charges, heat currents, or external fields,  the most recent strategy is to create devices that do not require the spin to charge conversion, but use the spin degree of freedom only to store and process information \cite{Datta,Khach,Menzel}. This idea is very appealing, because of the variety of systems, in which the micro- and nanoscopic magnetic dipoles with different anisotropy axes can be nowadays artificially produced. The list includes atomic spin ensembles \cite{Zhou,Menzel}, magnetic nanoarrays \cite{Datta,Mengotti,Marrows,Schumann,ZabelN}, structured multilayers and superlattices \cite{Cowburn:PRL2012,Zabel,Bedanta:PRB2006}, colloids \cite{Mikuszeit:PRB2009}, and molecular systems \cite{Sessoli:JACS2006,Wernsdorfer:NatMat2007}.
\begin{figure}
\includegraphics*[width=0.9 \columnwidth]{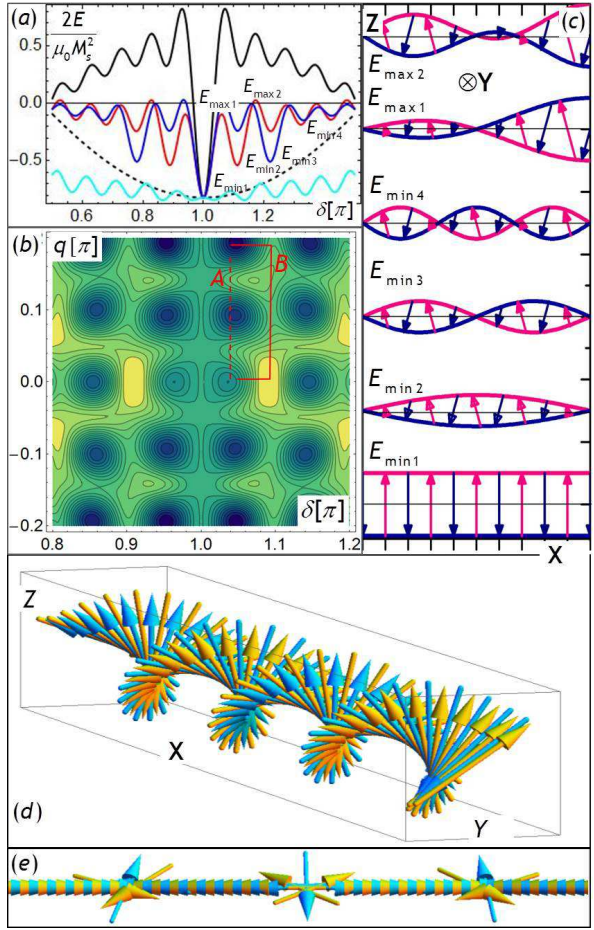}
\caption{(a,b) One- and two-dimensional representation of the energy of modulated helices in a dipolar chain consisting of ten moments as a function of $\delta$ and $q$; (c) $S_z(r)$ for several energy levels. The thick lines in (c) show the envelope lines of double-helices corresponding to the magnetic moments (arrows) in two sublattices. The dashed parabola in (a) corresponds to the energy of the harmonic spiral ${\bf S}(x)={\bf S}(\sin \delta x,0,\cos \delta x)$; the black, red, and blue curves correspond to $E(\delta)$ of modulated helices with $K$=0 and $q=0$, $q=\frac{\pi}{N}$, and $q=\frac{2\pi}{N}$ respectively, while the cyan line shows $E(\delta)$  for $q=\frac{\pi}{N}$ and $K=2.5 D$. The energy scale in (b) goes linearly from -0.8$D$ (blue) to +0.8$D$ (yellow); (d) Three-dimensional representation of an intermediate helix state found in the SD simulations for N=81 (video SI2) for $K\approx -1.23 \mu_0 M_s^2$ at $T<1$K; (e) End-configuration of SD-simulations for $K=0$ and identical to (d) starting configuration.}
	\label{fig:fig1}
	\end{figure}
In order to transmit and process information without electric currents or external fields intrinsic interactions are needed. The most ubiquitous in different kinds of magnetic systems interactions are the direct and indirect (e.g. Ruderman-Kittel-Kasuya-Yosida (RKKY)) exchange, and the dipolar coupling. Short, antiferromagnetically coupled via RKKY- or dipolar interaction chains ("spin leads"), have been already utilized to transmit the information of the state of an "input" ferromagnetic dot to the "gate" dot \cite{Khach,Cowburn:PRL2012}. It has been also realized that detailed structure of those antiferromagnetic (AF) states depend on the chain length \cite{Khach2}. For such nano-sized systems, however, there is up to now no clear understanding how the competition between the finite length of those chains and the order of interactions manifest itself in time-dependent magnetic order \cite{Khach2}. Especially intriguing is the aspect elaborated in \cite{Skomski:JAP2012}, showing that boundary conditions might induce topologically protected excitations. Clearly, this lack of knowledge hampers further development of the all-spin-based information technology.

In this paper, studying theoretically the size-dependence of magnetic order in chains with exchange, RKKY, and dipolar interactions, we discover a particularly simple law: magnetic states corresponding to the modulated helices with integer number of twists, that are commensurate with the chain's length, are topologically stable and correspond to local energy minima separated by energy maxima. The AF state, used so far for the spin leads, is a limiting, low energy case of the double helix with an infinite periodicity. With increasing number of twists the energy of the double helix increases but, once achieved, remains stable at small temperatures. The higher energy levels can be reached by rotating one of the chain ends like the winding up of spring driven clocks. In contrast to the clocks, a magnetic system may be "clicked into place" for integer number of twists.

We confirm these results using Monte-Carlo (MC) and Landau-Lifshitz-Gilbert Spin Dynamics (SD) simulations and a magneto-mechanical model. We also demonstrate that a chain can be forced towards one of the topologically stable states by rotation of end-spins. By further rotation, the helix can be winded up towards higher energy levels and pinned in this stable state to store energy. Unpinning this spin at a later time leads to release of the stored energy, which can be used to perform work or transfer information. The proposed device can be realized in structured multilayers \cite{Cowburn:PRL2012,Zabel}, chains of nanoparticles or molecules \cite{Sessoli:JACS2006,Wernsdorfer:NatMat2007}, nanoarrays \cite{ZabelN}, and atomic chains \cite{Khach}.

The starting point of our calculations is a linear (along $x$-axis) chain of $N$ dipoles coupled by either exchange or RKKY or dipolar interaction and possessing a uniaxial (easy axis or easy plane) anisotropy $K$ arising due to the magnetocrystalline anisotropy, particle shape, or higher order multipolar contributions. Here we describe results for the most complicated dipolar coupling. However, they can be straightforwardly adopted to the RKKY or exchange interactions.
\begin{equation}\label{eq:1}
H = D\sum_{ij}(\frac{\mathbf S_{i}\\
\mathbf\cdot\mathbf S_{j}}{r_{ij}^{3}}-3\frac{(\mathbf
S_{i}\mathbf\cdot \mathbf r_{ij})(\mathbf S_{j}\mathbf\cdot\mathbf
r_{ij})}{r_{ij}^{5}})-K\sum\limits_{i}(\mathbf{S}_{i}^{x})^2
\end{equation}
where $\mathbf{S}_{i}$ is a three-dimensional unit vector and $\mathbf r_{ij}$ denotes the distance-vector between moments $i$ and $j$. All information about the saturation magnetization $M_s$ of a particle is as usual hidden in the interaction constant $D=\mu_0 M_s^2/2$. First, we look for the exact total number of critical points of Hamiltonian (\ref{eq:1}) solving the set of equations $\vec\nabla {\sum\limits _{i,j}^N} H=\vec 0$ as explained in SI1 \cite{SI1}. We reveal $3\cdot 2^N$ stationary points (minima, maxima, or saddle points) related to non-collinear solutions and the question is how the magnetic configurations corresponding to these metastable states look like.

To answer this question analytically we utilize the method used for construction of spin-density waves. The magnetic structure is regarded as a superposition of spirals in the 2D Brillouin zone under the requirement of the constant magnetic moment at all sites. The energy of the constructed structure is then analyzed. The spin-spirals can be described via vectors $\delta$ and $q$ in the form (i) ${\bf S}(r)={\bf S}[\sin(\delta r)\cos (q r),\sin(\delta r)\sin (q r),\cos(\delta r)]$ or (ii) ${\bf S}(r)={\bf S}[\sin(\delta r),\cos(\delta r)\sin (q r),\cos(\delta r)\cos (q r)]$. For $q=0$ and ${\textbf{r}}\parallel {\textbf{Ox}}$, for example, (i) gives a spin-spiral ${\bf S}(x)={\bf S}[\sin(\delta x),0,\cos(\delta x)]$ in the $xz$-plane. The energy of this spin configuration as a function of $\delta$ is plotted as dashed line in Fig.~\ref{fig:fig1}(a). It is seen that the energy is minimal ($E_{min1}$) at $\delta =\pi$; i.e., corresponds to the antiferromagnetic (AF) alignment of neighboring spins and is known as the ground state of a chain with easy $xz$ plane (see Fig.~\ref{fig:fig1}(c)). There is only one minimum in this case and the envelope lines of AF structure Fig.~\ref{fig:fig1}(c) are straight. If, however, $({\delta,q})$ differ from these special values the energy spectrum changes dramatically: it adopts many local energy minima and maxima in good agreement with our analytical analysis. The contour plot of the energy $E(\delta,q)$ for a chain of $N=10$ dipoles is shown in Fig.~\ref{fig:fig1}(b). Several cross-sections of this two-dimensional energy surface for different $K$ are exemplified in Fig.~\ref{fig:fig1}(a). Fascinatingly, the minima occur for all $\{q,\delta\}=\{\pm\frac{n\pi}{N},\pi\pm \frac{m\pi}{N}\}$ for $K<-2.3D$ and $\{q,\delta\}=\{\pm\frac{n\pi}{N},\pi\pm(\frac{\pi}{2N}+ \frac{m\pi}{N})\}$ for $K>2.3D$ with integer $m\in [0,N]$ and $n\in [0,N]$. In other words, the energy spectrum becomes discrete and the energy gaps become size dependent; i.e., we observe all components of quantum confinement. The deepest minima; i.e., stablest energy levels correspond to magnetic helices (MH) with integer number of turns along the chain as shown in Fig.~\ref{fig:fig1}(c) and can be uniquely described using quantum numbers $m$ and $n$ defining the two characteristic, commensurate to the chain length wave vectors $\delta$ and $q$. The number of helix-turns is given by the smaller wave-vector. For example if $\delta\geq \pi$ then $q=\pi/N$ corresponds to one helix-turn of the entire chain.

\begin{figure}
\includegraphics*[width=0.9 \columnwidth]{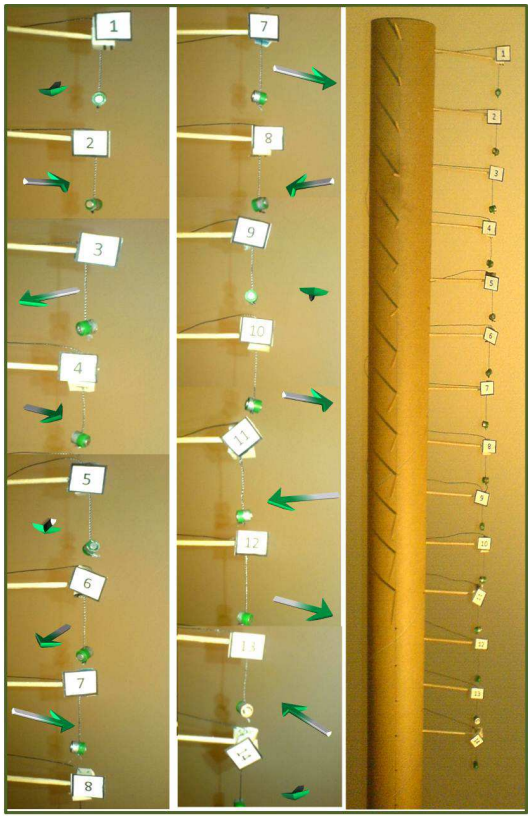}
\caption{Magneto-mechanical model \cite{SI1}. The magnets do not contain any connecting wire. They are hanging each on its own, soft non-magnetic filament and coupled via magnetostatics only.  The hole is larger than the diameter of the filament and magnets may rotate without screwing it.  Left: stable spin helix with modulation vector $q=\frac{3}{2}\pi/14$. Right: general view of the model. Silver and green colors represent the north and the south poles respectively. }
\label{fig:fig2}
	\end{figure}

To explore complete configurational space beyond the two-wave-vectors approximation MC and SD simulations have been performed. Additionally, a magneto-mechanical model presented in Fig.~\ref{fig:fig2} and described in \cite{SI1} has been constructed. While the MC procedure has been developed to describe the equilibrium properties of many-body systems, the SD gives an exact dynamical path from one configuration to another. The details of the numerical procedure are described in \cite{Vedmedenko:CPC,Nowak,SI1}. According to the MC analysis the stable equilibrium configuration of a dipolar chain of length $N$ for $kT>0.1 D$ and $-1.2 D<K<2 D$ indeed corresponds to $E_{min2}$ in Fig.~\ref{fig:fig1}(c); i.e., to $(q,\delta)=(\pi/N,\pi\pm\frac{\pi}{N})$. Only at lower temperatures or higher anisotropy the ground AF state can be achieved.
Our SD simulations demonstrate that the exact dynamical path towards the global energy minimum strongly depends on the starting configuration and passes through several energy MH states predicted above. Starting with the $ S[\sin((\pi+\frac{\pi}{N}) x),\cos((\pi+\frac{\pi}{N}) x),0]$  state, e.g., one ends up with a half-turn helix for anisotropy comparable with the dipolar energy. Another example of dynamical relaxation is recorded in video SI2 \cite{SI1}, while one of intermediate configurations is shown in Fig.~\ref{fig:fig1}(d). Even for $K=0$ the system often freezes in modulated structure shown in Fig.~\ref{fig:fig1}(e). In accordance with the simulations, the most probable stable states of the experimental chain established after mechanical agitation were spin spirals with $\pi$ or $\pi/2$ twist. Additionally to the easy-plane rotation of magnetization, a systematic modulation of the easy-axis component was also clearly observable. The less probable but still stable spiral with $q=\frac{3}{2}\pi/N$ configuration is documented in Fig.~\ref{fig:fig2}.

\begin{figure}
\includegraphics*[width=0.9 \columnwidth]{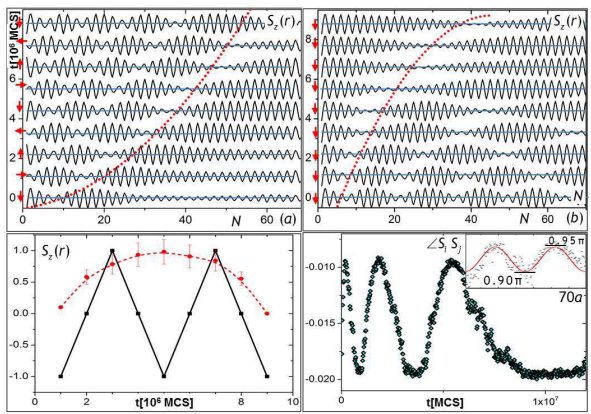}
\caption{MC simulations of a linear chain of 70 dipoles with $K\approx -1.23 \mu_0 M_s^2$ at $kT<1$K. (a) Winding up of the spiral. Each snapshot of $S_z(r)$ corresponds to a relaxed state. The very left moment (red) has been rotated with velocity of $(\pi/2)/2\cdot 10^{5}$ MCS; (b) Time dependence of the very left (black line) and the very right (red line) moments; (c) Release of the stored energy: the left moment is fixed downwards; (d) Non-linear time dependence of the chain-averaged $\langle S_z\rangle$. A typical angle distribution in a relaxed helix is shown as inset.}
	\label{fig:fig3}
	\end{figure}

The reasons for the stability of MHs are following: (i) the internal fields in these structures coincide with the orientation of dipoles, and (ii) those states cannot be transformed into the ground state by continuous deformation; i.e., they are topologically stable. Hence, if one fixes the ends of a chain, the helical structure will remain at the local energy minimum. Next, we want to use this property to achieve different $(\delta,q)$ states artificially. To do so we propose to rotate magnetization of one of the terminal dipoles introducing energy into the chain. The consequences of this procedure are documented in MC simulations (Fig.~\ref{fig:fig3}), SD simulations (videos SI3,4,6), and magnetomechanical model (Video SI5).

The Monte-Carlo simulations have been performed for chains of length $N=70a$. The equilibrium winding up process is shown in Fig.~\ref{fig:fig3}(a). Initially, the slow annealing procedure has been applied. At the end of relaxation, at $kT=0.05 D$, the chain has adopted the MH state with $q=\pi/N$; i.e., the entire chain acquired a half of a modulation period. This state was taken as the start configuration for the winding up process. Next, the first spin has been rotated with the velocity $\frac{\pi/2}{2\cdot 10^5{\rm MCS}}$, because our analysis showed that the period of $2\cdot 10^5$ MCS was long enough to achieve a new equilibrium state. After nine subsequent rotations, which are illustrated Fig.~\ref{fig:fig3}(a), the chain arrived at the stable $q=5\pi/2N$ helix. The other end of the chain remained free. In Fig.~\ref{fig:fig3}(b) the time dependence of the z-component of magnetization for the first and the last spins are monitored. One sees that the rotation of the last spin is somewhat delayed comparably to the rotation of the first spin because of the system's retarding in the stable energy levels. Another interesting observation concerns the propagation velocity of a knot in the spiral. The velocity of propagation decreases with increasing $q$ as shown by dashed red line and can be described by a function of the form $v=at^2$ with the negative acceleration $a$. The backward process is depicted in Fig.~\ref{fig:fig3}(c),(d). The first spin was fixed now and the chain's relaxation was monitored. The process was mirrored and the spiral was wound down performing work. The rotation was again decelerated as shown in Fig.~\ref{fig:fig3}(d). The whole process, however, took larger time as the spiral has been released at the local energy minimum.

The MC results are in very good agreement with SD simulations shown in video material SI3,4 for dipolar coupling and SI6 for ferromagnetic exchange interaction. The time scale depends on the strength of magnetic interactions and damping parameters. For a chain made out of 35$\times$35$\times$2 nm nanoparticles, interparticle distance 20 nm, and Gilbert damping of 0.01, the entire winding up process of video SI3,4 corresponds to milliseconds. Additionally, SD reveals new interesting aspects of the winding process. The first, evident conclusion is that the velocity of the entire process depends on the velocity of forced rotation of the first spin. The unexpected finding is that formation of each new turn of the magnetic helix is accompanied by an abrupt change in the magnetization of several dipoles in case of dipolar interaction. Initially, the neighbors of the first dipole rotate coherently with a certain phase shift. The internal energy of the chain climbs thereby towards a local energy maximum (e.g. $E_{max1}$ in Fig.~\ref{fig:fig1}). When the rotation angle approaches a critical value, the local maximum is arrived, and an abrupt change in the orientations of dipoles happens as the system falls down into the next stable state corresponding to two turns of the helix (see video SI3 \cite{SI1}). During this process part of the dipoles continue to rotate in the direction of forced winding up, while another part rotates in the opposite direction. In our magneto-mechanical experiment, the states with larger $q$'s are also achieved after the magnetization jumps described above, as demonstrated in video SI5. Hence, magnetic helix can be "clicked into place", what is different from Dzyaloshinskii-Moriya spin spirals \cite{Menzel} as discussed in \cite{SI1}. The "click into place" behavior during the winding up exists also for $q=0$, i.e. for $S_x=0$.
If we repeat the same procedure for a chain coupled via nearest-neighbor ferromagnetic exchange interaction with the same interaction strength and damping, the behaviour is the same but no jumps occur (see video SI6\cite{SI1}), because the relaxation time of the chain becomes comparable with the angular rotation velocity. When a wound up chain is released (video SI4), the stored energy becomes large enough to overcome several barriers.
The unusual properties of MH open several interesting perspectives in view of applications. The probably most important one concerns the new way of energy storage. One of the oldest but still actively utilized in a number of applications methods is the spring windup technique. An example, familiar to everyone is a clockwork device mechanically powered by a mainspring. In this method one end of a mechanical spring is fixed, while the other is rotated until the spring is wound up. Then the latter spring terminal is released and the stored potential energy transforms into the kinetic energy of a clock arrow, a motor, a pump etc. Similar procedure can be applied to the MH's. By rotation of one chain's end via local fields or spin-polarized currents the helix can be forced towards smaller periodicities and, thus, higher energy. The system can be leaved in this new stable configuration to store introduced energy. The stored energy can be then transformed into its mechanical or magnetic counterparts to make work as visualized in videos \cite{SI1} and Fig.~\ref{fig:fig3}.

In conclusion, the results described above demonstrate that the finite magnetic chains coupled by a exchange, RKKY or dipolar interactions possess a quantized energy spectrum depending on the chain geometry, material of elements, and the shape of particles. This unique energy spectrum results in topologically protected configurations in form of magnetic helices with integer number of revolutions, explains recent experiments \cite{Bedanta:PRB2006} and opens broad perspectives for future investigations concerning the dynamics of this nontrivial system and quantum effects on a length scale of a few {\AA}.

{\em Acknowledgements.}
Support by the DFG (SFB 668) and Landesexcellenzinitiative Hamburg (LEXI) is gratefully acknowledged.


\begin{thebibliography}{99}

\bibitem{Khach}
A. A. Khajetoorians, J. Wiebe, B. Chilian, and R. Wiesendanger, Science \textbf{332}, 1062 (2011).
\bibitem{Skomski:JAP2012}
R. Skomski et al., J. Appl. Phys. \textbf{111}, 07E116 (2012).
\bibitem{Zhou}
L. Zhou et al., Nature Phys. \textbf{6}, 187 (2010).
\bibitem{Menzel} M. Menzel et al., Phys. Rev. Lett. \textbf{108} 197204 (2012).
\bibitem{Datta}
B. Behin-Aein, D. Datta, S. Salahuddin, and S. Datta, Nature Nano, \textbf{5}, 266 (2010).
\bibitem{Khach2}
A. A. Khajetoorians et al., Nature Phys. \textbf{8}, 497 (2012).
\bibitem{Mengotti}
E. Mengotti et al., Nat. Phys. \textbf{7}, 68 (2011).
\bibitem{Marrows}
J. P. Morgan, A. Stein, S. Langridge, and C. H. Marrows, Nat. Phys. \textbf{7}, 75 (2011).
\bibitem{Schumann}
A. Schumann, P. Szary, E. Y. Vedmedenko, and H. Zabel, New J. Phys. \textbf{14}, 035015 (2012).
\bibitem{ZabelN} M. Ewerlin et al., Phys. Rev. Lett. \textbf{110}, 177209 (2013).
\bibitem{Cowburn:PRL2012}
A. Fernandez-Pacheco et al., Phys. Rev. B \textbf{86}, 104422(2012).
\bibitem{Zabel}
M. van Kampen et al., J. Phys.: Condens. Matter \textbf{17}, L27 (2005).
\bibitem{Bedanta:PRB2006} S. Bedanta et al. Phys. Rev. B \textbf{74}, 054426 (2006).
\bibitem{Mikuszeit:PRB2009}
N. Mikuszeit et al., Phys. Rev. B \textbf{80}, 014402 (2009).
\bibitem{Sessoli:JACS2006}
K. Bernot et al., J. Am. Chem. Soc. \textbf{128}, 7947 (2006).
\bibitem{Wernsdorfer:NatMat2007}
W. Wernsdorfer, Nature Mat. \textbf{6}, 174 (2007).
\bibitem{Vedmedenko:CPC}
E. Y. Vedmedenko and N. Mikuszeit, ChemPhysChem \textbf{9}, 1222 (2008).
\bibitem{Nowak}
U. Nowak, Ann. Rev. of Comp. Phys. IX, 105 (2001).
\bibitem{SI1}
See Supplemental Material SI1-SI6 at http://link.aps.org.

\end{thebibliography}
\end{document}